# Search algorithms as a framework for the optimization of drug combinations

Calzolari D. [1,3], Bruschi S. [1,3], Coquin L. [1], Schofield J. [1], Feala J. [2], Reed J.C. [1], McCulloch A.D.[2], Paternostro G. [1, 2]

[1] Burnham Institute for Medical Research

10901 North Torrey Pines Road

La Jolla, CA, 92037

[2] Department of Bioengineering

University of California, San Diego, CA 92093

[3] These authors contributed equally

**Corresponding author:**

Giovanni Paternostro
Burnham Institute for Medical Research
10901 North Torrey Pines Road
La Jolla, CA, 92037
Tel. (858) 713 6294
Fax (858) 713 6281
e-mail: giovanni@burnham.org

ABSTRACT

Combination therapies are often needed for effective clinical outcomes in the management of complex diseases, but presently they are generally based on empirical clinical experience.

Here we suggest a novel application of search algorithms – originally developed for digital communication – modified to optimize combinations of therapeutic interventions. In biological experiments measuring the restoration of the decline with age in heart function and exercise capacity in *Drosophila melanogaster*, we found that search algorithms correctly identified optimal combinations of four drugs with only one third of the tests performed in a fully factorial search. In experiments identifying combinations of three doses of up to six drugs for selective killing of human cancer cells, search algorithms resulted in a highly significant enrichment of selective combinations compared with random searches. In simulations using a network model of cell death, we found that the search algorithms identified the optimal combinations of 6-9 interventions in 80-90% of tests, compared with 15-30% for an equivalent random search.

These findings suggest that modified search algorithms from information theory have the potential to enhance the discovery of novel therapeutic drug combinations. This report also helps to frame a biomedical problem that will benefit from an interdisciplinary effort and suggests a general strategy for its solution.

AUTHOR SUMMARY

This work describes methods that allow the identification of drug combinations that might alleviate the suffering caused by complex diseases. Our biological model systems are the physiological decline associated with aging and selective killing of cancer cells. The novelty of this approach is based on a new application of modified methods derived from the field of digital communications.

**INTRODUCTION**

The problem of combination therapy has medical and algorithmic aspects. Medically, we are still not able to provide effective cures for most chronic, complex diseases that are the main causes of death and disability, nor are we able to address the progressive age-related decline in human functional capacity. Algorithmically, when biological dysfunction involves complex biological networks, therapeutic interventions on multiple targets are likely to be required. Because the effect of drugs depends on the dose, several doses need to be studied, and the number of possible combinations rises quickly. For example, many cancer chemotherapy regimens are composed of 6 or more drugs from a pool of more than 100 clinically used anticancer drugs and exploring even larger combinations might be justified [1]. If we were to study all combinations of 6 out of 100 compounds (including partial combinations containing only some of these compounds) at 3 different doses we would have $8.9 \times 10^{11}$ possibilities. This example shows that the problem requires a qualitatively new approach rather than simply more efficient screening technology.

Combined drug interventions are a common therapeutic strategy for complex diseases such as hypertension and cancer. As pointed out recently for cancer therapy [2], most therapies were initially developed as effective single agents and only later combined clinically. A possible approach to the exploration of new therapeutic activities not present in individual drugs is based on the exhaustive study of all possible combinations of pairs of compounds [3]. This "brute force" approach has detected many interesting novel pairs of compounds [3], but the resulting exponential expansion in the number of possibilities precludes the comprehensive exploration of larger combinations.

Fitzgerald *et al.* [4] have recently argued that the future of combination therapy lies in the development of accurate quantitative network models that capture the mechanistic interactions of cellular and organism physiology. The authors acknowledge that we do not yet know what these models will look like, and that systems biology research is still data-limited for this purpose. Indeed their recent review does not report any successful application of this approach to combination therapies.

Here we suggest a novel solution to the problem of combination drug therapy, making use of search algorithms originally developed for digital communication. When modified in several key aspects, these search strategies can be used to find more effective combined therapeutic interventions without the need for a fully factorial experimental design (testing all possible combinations of drugs for all selected doses). These algorithms may also provide a framework upon which information from system-wide molecular data (e.g. transcriptomics and metabolomics) and from mechanistic computational networks models can be superimposed.

*Rationale for the suggested algorithms*

To understand the motivation for our work it is important to consider that, even if simulations might play a role, the intended use of the algorithms is not entirely *in silico*, but partially *in vivo* or *in vitro*, using high-throughput biological measurements in organisms or isolated cells, respectively. This approach becomes increasingly relevant because high-throughput measurement technology, initially developed by drug companies for the screening of large libraries of compounds in multi-well plate formats, is now more and more available to the scientific community.

It is useful to regard the information processing by our experimental systems as parallel biological computations, since the algorithms we are using are indeed derived from algorithms that were implemented *in silico* in other scientific fields. Parallel measurements are suitable for multi-well high-throughput technology.

There are requirements regarding the computational complexity of the algorithms that limit the choice of suitable approaches. These requirements are discussed in more detail in section 2.4 of Results. Both the number of operations and computational costs unique to *in vivo/in vitro* algorithms should be considered.

Algorithm design requires the application of an appropriate structure to the data. Although there are many options to represent the space of possible drug combinations, we used a tree representation with drug combinations as nodes linking to all possible additions of one drug in the next level. Individual drugs form the base of the tree and combinations of maximum size are at the top (Figure 2). When exploring the drug combination tree going from smaller to larger combinations, as in the algorithms we suggest, we are giving more weight to lower-order drug interactions. This is consistent with data available on adverse drug interactions, which are reported mostly for two-drug combinations [5,6]. Estimating the optimal size of a combination is a different problem, examined in detail in the Discussion. The beneficial effect of a combination is also due to additive components (not depending on interactions) and to multiple higher-order effects.

The search algorithms we suggest are derived from sequential decoding algorithms. These were chosen in part because of similarities among the data trees to be searched in the biological and decoding applications (Figure 2). Sequential decoding algorithms are used for convolutional codes, in which nearby nodes in the data tree are related, similarly to different but partially overlapping combinations of drugs.

Another feature of sequential algorithms that fit our purposes is the use of a list-based memory of the path taken to reach each node. We provide in the Discussion a detailed argument suggesting that a suitable algorithm should be able to integrate all available information on the state of the system with that obtained by iterative measurements. The integration should take place at every iteration within the algorithm, rather than being a weighted average of different methods applied separately. The presence of the updated list as a guide for each iteration provides our algorithms with a natural mean of information integration.

Both the fully factorial dataset we show in Figure 1 and the complex structure of the biological networks that are being reconstructed in systems biology supports this expectation of frequent non-linearities in phenotype measurements along the data tree. Therefore we are interested in algorithms that can search within a solution space presenting substantial non-linearities. If the relation among drugs in a combination were linear, the best algorithm would simply determine the best dose in single drug measurements and use these to obtain the best combination. If on the contrary non-linearities were extreme, the use of stochastic algorithms might be preferable. Stochastic algorithms (see also Discussion) can cope with multiple local minima in the solution space, but they do so by incorporating a random element. This requires a price in terms of computational cost, and the performance of stochastic algorithms is therefore often not as good as that of more tailored algorithms [7,8]. The algorithms we suggest can cope with moderate and variable non-linearities by going back to previous nodes in the tree.

Starting with the stack sequential algorithm, which was developed to search for optimal decoding in the field of digital communications [9], we describe and test algorithms that can be used to search for an optimal combination of a sizeable number of drugs, by testing only a small subset of all possible combinations. The algorithms are useful for large combinations, where collecting fully factorial datasets is not feasible. We present results obtained from simulations in a computational model of cell death and from experiments using two models with complementary biological properties: (i) restoring the decline with age in heart function and exercise capacity in *Drosophila melanogaster;*and (ii) selective killing of human cancer cells.

The first *in vivo* experimental model has the advantage of including the complexity of whole organism interventions, while the second *in vitro* model has the potential for markedly higher throughput testing. These models are also representative of two different general types of multi-drug interventions: one type aims at improving function, while the other is based on the induction of cell death, a selective disruption of network function. Results suggest that optimal or near-optimal combinations of compounds can be found in these systems with only a small fraction of the number of tests as a fully factorial design, and with significantly higher efficacy than random searching. In summary the contributions of this work are:
- Constructing a novel problem statement for the search of drug combinations, using a novel approach to systems biology (see also Figure S1 in Supplementary material).
- Collecting exhaustive experimental measurements (the fully factorial dataset) sufficient to solve the problem conclusively.
- Constructing a computational method to solve the problem approximately with fewer experimental measurements (the search algorithms). The suggested algorithms are modeled on algorithms already used in other fields, our main original contribution is in their novel application.
    - Providing additional experiments to support the generality of the approach.

**RESULTS**

1. A FULLY FACTORIAL DATASET OBTAINED IN DROSOPHILA
A fully factorial dataset is a dataset where all possible combinations of drugs for the selected doses are tested. The dataset was obtained from biological measurements in a living organism, *Drosophila melanogaster* (the fruitfly).

A detailed account of the *Drosophila* cardiac aging model was presented previously [10]. We performed an initial screen of compounds for their effects on cardiac aging in *Drosophila*, selected for their general effects on multiple biological functions, previously demonstrated low toxicity and, for some compunds, known effects on aging in other models.

After screening 44 compounds individually at multiple doses (a total of 300 groups, each composed of 10-20 flies), we chose two doses each of four compounds for more comprehensive measurements of their combined effects on three age-related phenotypes: the declines in maximal heart rate, exercise capacity and survival. The selected compounds (see Methods for doses in the fly food) were: doxycycline, a broad spectrum antibiotic and inhibitor of mitochondrial protein synthesis [11]; sodium selenite, an essential trace mineral and cofactor of many metabolic enzymes; zinc sulfate, another trace mineral and cofactor of many metabolic enzymes; and resveratrol, a phenolic antioxidant with an action on proteins linked to aging [12].

The compounds were fed to flies from the age of 7 days to the age of 30 days. We have previously shown cardiac physiological changes with age in 30 day-old flies [10]. The maximal heart rate was measured at the age of 30 days. Climbing velocity was measured every 5 days between the ages of 15 to 30 days, using a non-invasive procedure. We studied 10 male flies for climbing and 10 female flies for the cardiac measurements. Survival to 30 days was also measured in these flies. Figure 1 illustrates the fully factorial dataset consisting of 81 combinations of 4 drugs, using 2 doses for each drug (1 control, 8 individual tests, 24 groups of 2 combined drugs, 32 groups of 3 combined drugs and 16 groups of 4 combined drugs).

The number on the right of each combination in Figure 1 is a summary score (z-score) obtained from the three phenotypes mentioned: the declines with age in maximal heart rate, climbing velocity and survival. Each value was normalized by dividing by a weekly control, then for each group subtracting the population mean and diving by the population standard deviation. The z-scores from the three phenotypes were then averaged to yield a summary z-score that equally weights each of the three measurements. Analysis of Figure 1 shows that, with a larger number of drugs in the combination, there is an statistically significant increase ($p<0.05$) in the percentage of treatments that have an improved z-score compared with untreated controls of the same age.

The landscape (see section in Discussion on control landscapes) obtained from this dataset has 7 local maxima and 1 global maximum in the phenotype z-score. The maxima correspond to drug-doses configurations for which the z- score decreases by changing any of the drugs by a single dose. We have also calculated the basin of attraction, i.e. the number of drug-doses configurations that will end up in a given maximum by following the maximal increase in z-score, and found that the global maximum corresponds to the largest basin. This is an example of how landscape terminology can be used to define moderate non-linearities suitable for the algorithmic approach we suggest.

2. THE ALGORITHMS

*2.1 Sequential decoding algorithms and the stack sequential algorithm*

In this section we introduce the drug combination optimization algorithms and show how they relate to the algorithms used in sequential decoding. Fully factorial datasets, where every possible drug combination is tested, grow exponentially with the number of drugs (n). See Supplement for the relevant equation and an example dataset (Table S1). In computational terms we say that the complexity is $O(a^n)$. The O-notation indicates the order of growth of an algorithm basic operation count as a function of the input size. An

Search algorithms for drug combinations

exponential growth is not practical for large n, therefore our aim is to find algorithms with improved efficiency, for example with a linear dependency on n, expressed as O(n).

The "stack sequential algorithm" was first proposed by Zigangirov and Jelinek for the sequential decoding of noisy digital signals [9,13]. As pointed out by Johannesson and Zigangirov [14], the word "stack" is used instead of the proper word "list" only for historical reasons. It is the most basic and simplest to describe of the sequential decoding algorithms.

The problem of finding the optimal estimate of the encoded sequence is described as a walk through a tree. To appreciate the analogy with the search for the optimal drug combination, the tree shown in Figure 2 can be compared with the trees used in one of the original descriptions of the stack sequential algorithm [13].

The stack is a sorted list of all examined combinations (best on top). The description of the stack sequential algorithm of the Jelinek paper [13] corresponds to the following adaptation to our problem:

S1 - At the beginning of the process the list contains only the measurement in the absence of any drug (the root of the tree of Figure 2)
S2 - The search is extended from the top of the sorted list. An extension corresponds to moving up one level in one of the branches of Figure 2. Combinations already used are ignored for future extensions.
S3 - The search ends when a combination of maximum size is reached. This is equivalent to reaching the top of the tree of Figure 2.

Since we are looking for the best combination, and not for the best path, in our case we do not delete any measured combination from the list. Instead, when a combination has already been used for extension in the tree, we move to the next combination in the sorted list. As indicated in Figure 2, we do not combine different doses of the same drug with each other, to limit the size of the search, but this is not an essential feature.

This algorithm is effective in searching combinations where the effect is not purely additive, because it can overcome non-linearities by going back to previous nodes in the tree.

*2.2 Three classes of algorithms for searching the data tree*

A family of related algorithms can be derived from the basic structure of the stack sequential algorithm described in the previous section, adapted to different requirements, and with different trade-offs between complexity and performance. This is similar to the case of sequential decoding. Examples of other algorithms that are part of the sequential decoding family, with trade-offs partially analogous to those we have implemented, are the Fano algorithm and the Creeper algorithm [14].

This family of algorithms can be divided into three basic classes that differ in their approach to the data tree of Figure 2. Figure 4 shows the structure of these three classes. The class that follows the same direction of search within the data tree as the stack sequential we call SS; the class that searches the tree in the other direction, from the top down, we call TD; and the class combining both approaches (starting as SS and continuing as TD) we call SS-TD (see Figure 4). Below we describe the implementation of each class used in this paper, which we call SS', TD' and SS-TD'.

*2.3 Notation*

Let *DRUGS* represent the set of all drugs under consideration, and let *DOSES* represent the set of all possible doses. Additionally, let *n* be the number of drugs and *m* the number of possible doses.

Let a drug be denoted by *D* while a dose is represented by *d*. The ordered pair (*D*, *d*) represents the drug along with its dose. Let C represent a collection of drug-dose pairs (a drug combination).

Let function *Score*(*C*) assign a score *z* to the drug combination C and save (C,z). Let C_LIST = [($C_1$ $z_1$), ($C_2$ $z_2$), … ] represent a list of drug combination-score pairs. For any collection $C_i$, we refer to the cardinality of the collection $|C_i|$ as size (size of the combination).

The problem of selecting $C_{opt}$, the optimal drug combination that maximizes *Score*(C) is a combinatorial optimization problem for which we propose a number of algorithms.

It is important to note that *Score*(C) is the only step that is not executed *in silico* but is measured *in vivo* or *in vitro* (biologically). A ranked summary list with all the measured combinations is obtained at the end of the procedure.

*2.4 SS' algorithm*

This algorithm starts by evaluating all individual drug-dose pairs, and then incrementally adds (*D*, *d*) pairs extending from the pair producing the maximal benefit. If at the *i*-th step, the addition does not increase the benefit, the algorithm backtracks to choose the next most beneficial combination from the (*i*-1)-th step. The informal steps (S) of the algorithm, which are also presented in pseudocode in Figure 5, are as follows:

S1  Evaluate all drugs individually at all doses and rank them according to the biological score.
S2  Save only the best dose for each drug in the single drug list.
S3  Extract the best single drug and call it $C_{best}$.
S4  Combine $C_{best}$ with all other drugs (for all doses), increasing the size of the combination by 1 drug, measure the biological scores, and save the list of combinations of this size. At this step the algorithm moves one level upwards in the tree of Figure 2.
S5  If the score of one of the new combinations is better than $C_{best}$ use this combination as the new $C_{best}$ and return to S4.
     If none of the new combinations is better than $C_{best}$ backtrack to the next best combination in the list of the previous size, call it $C_{best}$ and return to S4.
S6  Limit the number of backtracks to a specified value.
S7  Repeat S4 to S6 until the maximum size for the combinations is reached.

In this implementation we introduced two features that make the algorithm more appropriate for our application.

In S2 we choose only the best dose to extend from for the individual drugs. This is because if we need to return to this level after having combined the best single drug with all the others, it means that all lower order interactions (that is couples) were not beneficial, and therefore we prefer to change drug rather than trying a different dose of the same drug.

In S6 we limit the number of backtracks to limit the cost of the algorithm. We used a limit of 2 in all the experiments presented. This limit can be increased if the throughput of the technology we use for the biological measurements allows it. If we wish to increase this limit, we can make a choice among possible implementations that either backtrack only one level at a time or jump to any level that ranks next in the summary list. These implementations would have different complexity.

While SS' moves up in the data tree the number of measured combinations declines (see Supplement, Table S2). This algorithm therefore gives greater weight to lower order combinations in deciding which branches of the data tree we should explore. This is consistent with the expectation that lower order interactions among drugs are likely to be stronger than higher order interactions, as mentioned in the Introduction.

The experimental complexity of this implementation (both best and worst case) grows as $O(n^2)$ for the number of drugs, and $O(m)$ for the number of doses (see Supplement, legend to Table S2). Increasing the backtracking limit we might reach, in the worst case, the same complexity of the fully factorial, that is $O(a^n)$.

For algorithms including an *in vivo* or *in vitro* (biological) step we also have to consider other types of computational complexity, beside the number of operations. Biological measurements can take a very long time compared to any *in silico* step (several weeks are required for the Drosophila experiments) and may be limited by sample availability. This type of cost needs to be calculated separately for each application. In these algorithms there are also iterated cycles composed of biological measurements that can be done in parallel. All combinations formed in S4 above, extending from the best scoring combination, can be measured in parallel. Parallelization suits existing screening technology (e.g. multi-well plate robotics) but the number of cycles can also be limiting, again depending on the specific biological application.

*2.5 TD' algorithm*

The rationale motivating the development of top-down searches within the data tree is based on both the higher scores for larger combinations shown by the Drosophila fully factorial dataset of Figure 1 and the reduced number of combinations of higher order in all fully factorial datasets of this type, shown in Table S1 in the Supplement. These two factors led us to expect a higher probability of finding desired scores within combinations of larger size, and supported the development of algorithms with a higher proportion of measurements in this region of the data tree. We can also easily modify the algorithms to stop once a combination with the desired score is found, and therefore we wish to increase the probability of finding these combinations early in the search.

The first steps are the same as those of the SS' algorithm (see Figure 6 for pseudocode):

**S1**  Evaluate all drugs individually at all doses and rank them according to the biological score.
**S2**  Extract the best single drug and call it $C_{best}$.

After the individual measurements the TD' jumps to search within combinations with the largest size and moves progressively down the data tree from there:

**S3**  Combine $C_{best}$ with all other drugs (for all doses) for combinations of maximum size, measure the biological scores, and save the ranked list of combinations. At this step the algorithm moves to the highest level in the tree of Figure 2.
**S4**  Save the list of combinations that score above the median.
**S5**  Count the occurrences of all drug-dose pairs and save them in a new list.
**S6**  Save only the most frequent dose for each drug in the new list.
**S7**  Create all possible combinations of the next smaller size, using the drugs in the list of most frequently occurring drug-dose pairs, and measure the biological scores. At this step the algorithm moves down one level in the tree of Figure 2.
**S8**  Return to S4 until reaching size 2.

The complexity of this algorithm is $O(a^n)$, and is therefore suitable only for searches within a small number of drugs. It is described only to make the construction of the next algorithm clear.

*2.6 SS-TD' algorithm*

This algorithm aims to combine the desirable features of the two preceding algorithms. It starts as a SS' up to combinations of J drugs and then jumps to the largest size combinations like the TD'. See Figure 7 for pseudocode.

The computational cost is limited by design, because we choose J so that the cost is always within 10% of the corresponding SS'; therefore the SS-TD' has the same complexity as SS', $O(n^2)$.

3. TESTING THE ALGORITHMS IN THE DROSOPHILA DATASET (IN VIVO)

The fully factorial dataset of Figure 1 was used to test the SS' and SS-TD' algorithms. Both algorithms were successful in finding the best combination (and 3 of the 5 best combinations) with a lower cost compared to an exhaustive search (24 and 27 tests out of 81 for the the SS' and SS-TD' respectively).

4. TESTING THE ALGORITHMS WITH COMPUTATIONAL SIMULATIONS ON THE APOPTOSIS NETWORK (IN SILICO)

We performed computational simulations of multiple interventions on the apoptosis network using the two algorithms described above . The computational model is based on the apoptosis network, hsa04210, of the KEGG database (www.genome.jp/kegg/). We Search algorithms for drug combinations11used the discrete apoptosis model described in our previous publication [1], where the discrete state of proteins at each node is determined by the strength of a signal from the neighboring nodes according to a logarithmic rule. In this model, the final life/death signal is calculated following the signaling in the directed network up to a final output node. The effect of a drug on a given node is modeled by changing the activity on that node, and calculating the corresponding change in the output life/death node.

We simulated selective killing of cells caused by drugs acting on the apoptosis network. All possible interventions on 6, 7, 8 and 9 proteins, using 3 doses, were simulated. We used the dataset containing all possible interventions to study the efficacy for selective killing of the two algorithms (SS' and SS-TD'), compared with randomly selected combinations of the same size (see Figure 8).

Both algorithms were significantly more efficient than random tests (p<0.0001). The SS-TD' was clearly superior in the frequency of identification of the very best combination, but the SS' also performed well (Figure 8). If a purely additive strategy were the optimal one, the SS' would find it, with no backtracks. However, this does not seem to be the case. In the fully factorial tests, larger combinations of up to 9 interventions were more effective than single or two-drug interventions in finding the most selective solution (p<0.0001).

We also performed an alternative simulation changing a large number of parameters (see Methods section), to test the robustness of these findings, and were able to confirm the behavior shown in Figure 8.

As suggested by the number of top combinations found by random sampling in Figure 8, these fully factorial datasets contained multiple maxima. We investigated a different group of 30 fully factorial datasets (using 8-drug combinations) where maxima were very few (less than 0.05% of the total). Not surprisingly, in these simulations, random tests never found the top combinations. However top combinations were found in 30% of the tests by the SS' algorithm and in 80% of tests by the SS-TD' algorithm. Furthermore the distances of the best solutions found from the real maxima (expressed in % of the optimal value) were: 9.2 ± 1.4 (mean ± SEM) for random tests, 4.7 ± 1.2 for SS' and 0.3 ± 0.1 for SS-TD'. All differences between groups were statistically significant (p<0.01).

5. TESTING THE ALGORITHMS IN CANCER CELL LINES (IN VITRO)

Two lymphoma cell lines, RS 11846 and DoHH2, were used. These cell lines were chosen for the simplicity of the culture conditions, aiming to validate the method. Future tests will explore selectivity including also normal cells and cells with different tumorigenic potential. The number of viable cells was measured using a luminescence test for ATP (ATPlite, PerkinElmer). We used three different doses (for 60 hours) of six drugs affecting cell viability: Vincristine, Etoposide, Rituximab, Apogossypol, Dexamethasone and qVD-OPH. The first five drugs can induce cell death as individual interventions while the last is an inhibitor of cell death.

We compared the SS-TD' algorithm with random combinations. After 36 tests for each cell line using individual doses we measured 91 combinations using the SS-TD' algorithm and 107 randomly chosen ones (107 was the maximum theoretical number of tests required by the algorithm). The steps followed the order: couples, triplets, sextuplets, quintets, quartets. The SS-TD' selectivity (mean 21.3 ± 2.4 %) was markedly better than that in the random approach (mean 1.9 ± 2.5 %, p<0.0001). Furthermore none of the five used the discrete apoptosis model described in our previous publication [1], where the discrete state of proteins at each node is determined by the strength of a signal from the neighboring nodes according to a logarithmic rule. In this model, the final life/death signal is calculated following the signaling in the directed network up to a final output node. The effect of a drug on a given node is modeled by changing the activity on that node, and calculating the corresponding change in the output life/death node.

We simulated selective killing of cells caused by drugs acting on the apoptosis network. All possible interventions on 6, 7, 8 and 9 proteins, using 3 doses, were simulated. We used the dataset containing all possible interventions to study the efficacy for selective killing of the two algorithms (SS' and SS-TD'), compared with randomly selected combinations of the same size (see Figure 8).

Both algorithms were significantly more efficient than random tests (p<0.0001). The SS-TD' was clearly superior in the frequency of identification of the very best combination, but the SS' also performed well (Figure 8). If a purely additive strategy were the optimal one, the SS' would find it, with no backtracks. However, this does not seem to be the case. In the fully factorial tests, larger combinations of up to 9 interventions were more effective than single or two-drug interventions in finding the most selective solution (p<0.0001).

We also performed an alternative simulation changing a large number of parameters (see Methods section), to test the robustness of these findings, and were able to confirm the behavior shown in Figure 8.

As suggested by the number of top combinations found by random sampling in Figure 8, these fully factorial datasets contained multiple maxima. We investigated a different group of 30 fully factorial datasets (using 8-drug combinations) where maxima were very few (less than 0.05% of the total). Not surprisingly, in these simulations, random tests never found the top combinations. However top combinations were found in 30% of the tests by the SS' algorithm and in 80% of tests by the SS-TD' algorithm. Furthermore the distances of the best solutions found from the real maxima (expressed in % of the optimal value) were: 9.2 ± 1.4 (mean ± SEM) for random tests, 4.7 ± 1.2 for SS' and 0.3 ± 0.1 for SS-TD'. All differences between groups were statistically significant (p<0.01).

5. TESTING THE ALGORITHMS IN CANCER CELL LINES (IN VITRO)

Two lymphoma cell lines, RS 11846 and DoHH2, were used. These cell lines were chosen for the simplicity of the culture conditions, aiming to validate the method. Future tests will explore selectivity including also normal cells and cells with different tumorigenic potential. The number of viable cells was measured using a luminescence test for ATP (ATPlite, PerkinElmer). We used three different doses (for 60 hours) of six drugs affecting cell viability: Vincristine, Etoposide, Rituximab, Apogossypol, Dexamethasone and qVD-OPH. The first five drugs can induce cell death as individual interventions while the last is an inhibitor of cell death.

We compared the SS-TD' algorithm with random combinations. After 36 tests for each cell line using individual doses we measured 91 combinations using the SS-TD' algorithm and 107 randomly chosen ones (107 was the maximum theoretical number of tests required by the algorithm). The steps followed the order: couples, triplets, sextuplets, quintets, quartets. The SS-TD' selectivity (mean 21.3 ± 2.4 %) was markedly better than that in the random approach (mean 1.9 ± 2.5 %, p<0.0001). Furthermore none of the five most selective combinations could have been found with the traditional approach of combining only drugs that are cytotoxic individually, since these five combinations all contained qVD-OPH. The cancer cell results are shown in Figure 9.

**DISCUSSION**

It might be argued that each drug combination and biological system will require a different search algorithm and that there is no reason to expect universality. The results reported here, obtained with very diverse systems and compounds, suggest otherwise. Additional rationale supporting the existence of optimization algorithms with general applicability to biological networks is provided by the shared properties of these networks (such as a scale free distribution [15], robustness and evolvability [16]).

In future studies it will be desirable to develop formal methods to assist in the choice of the individual drugs to be considered by the algorithms, and to determine which doses to study. It is reasonable to consider several doses spanning the IC50 or EC50, but in large combinations we should expect to use lower doses than those common for the same drugs as single agents. At least initially the compounds more appropriate for use in the algorithms are FDA approved drugs and well-known supplements, for which the preferred dosage as single agents is already known.

*Alternative approaches*

Several concepts (e.g. synergy) have been developed in the past for the study of combinations of mainly two drugs [3,17,18]. Synergy is useful but it is not a necessary property for the optimal combination In any case, our algorithm objective (finding the best combination) includes the case where this optimal result is due to synergy.

A PubMed search for algorithm and "combination therapy" identified 101 papers. All the abstract and the papers that appeared relevant were reviewed. Most papers describe sets of clinical rules derived from clinical experience or from randomized clinical trials, relevant to combinations of 2-3 drugs, to be implemented by physicians. These approaches were called therapeutic, diagnostic, treatment, management or decision algorithms. A few papers [19-21] describe algorithms to be implemented *in silico* and providing guidance for some drug combinations of small size, using disease specific information. None of these papers described search algorithms suitable for partially *in vivo* or *in vitro* searches as those we describe.

A recent interesting paper [22] describes the use of stochastic algorithms for the search for optimal drug combinations. The methods described are not directly suitable for parallel biological measurement but stochastic methods, for example genetic algorithms, can certainly be adapted for this purpose.

*Size of drug combinations*

Several current cancer chemotherapy regimens are composed of combinations of 6 or more drugs. Examples, indicated by their acronyms and followed by the respective

Search algorithms for drug combinationsnumber of drugs, are: BEACOPP 7, ChlVPP/EVA 7, MACOP-B 6, ProMACE-CytaBOM 9, MOPPEBVCAD 10, m-BACOD 6 [23-26].

When the algorithms suggested here search within a pool of drugs the best combination found can be of any size. In other words when searching within all possible combinations of different doses of 10 drugs, it is possible that the best combination emerging might be composed of only 3 drugs, as for example in the Drosophila dataset of Figure 1.

An important question wether we can determine the maximum number of compounds that a combination should have. Our opinion is that such a maximum limit cannot be set as a general rule, based on the following considerations:

- Our algorithms can be used for combinations of any compound with biological activity, including not only drugs but also natural products. There are several dietary components, for example wine, that have been suggested to have, at certain doses, beneficial effect on human health [27,28]. These dietary components contain a large number of partially unknown different chemical compounds.
- Toxicity does not necessarily limit the size of a combination, as discussed in the safety section that follows.
- The complexity of many biological networks leads us to expect that only an intervention on a large number of nodes might allow us to optimize their function. Our knowledge of these networks is however still incomplete and no precise calculations are possible.

*Information theory and search algorithms for optimal drug combinations*

We can think of drug interventions as transmitting information to biological networks. When we search for optimal drug combination the efficiency of transmission of information (the domain of information theory) is important, and it is therefore not surprising that some modified algorithms from digital communications, which are used to efficiently decode signals in the presence of noise, might be applicable. There are, however, also several differences that require modifications to these algorithms.

Among the similarities with the digital communication applications of sequential decoding algorithms are the following: the partial exploration of a tree of possible solutions, the dependence of the score on the previous steps of the algorithm, the objective of maximizing the score and minimizing the cost, and the use of an ordered list to store the solutions.

Among the differences are the following: the partially different data structure to be explored and the related possibility of jumping to different parts of the tree and even ignoring some steps (for example SS-TD class algorithms are not used for decoding and are unlikely to be useful), and the tendency of the largest combinations to have higher scores. The computational cost is also partially different. For example memory is not a limiting factor but the number of tests and the time required by each step are.

*Safety*

We would also like to discuss the drug safety implications of the use of drug combinations in general and of our approach more specifically. Of the two main types of adverse drug events, type A adverse events represent the majority [29]. These are dose-related and arise from the pharmacological action of the drug [29]. Type A adverse events are not necessarily increased in combinations if we use reduced doses of each drug. Furthermore the objective metric of the algorithm can incorporate the reduction of adverse effects. An example is the choice of selective cell death for the cancer cell measurements we report, rather than just the killing of cancer cells. If we were to find a large therapeutic combination that had an extremely selective action only on cancer cells (or on a particular cancer), this would have a greatly improved safety profile compared to any of the existing chemotherapeutic regimens.

The second major type of adverse drug events, type B, is much more rare and not dose-related. These adverse events are at least in part genetic [29] and should be ameliorated by including genomic data as one of the omics components of our algorithms in future implementations. As for single drugs, medication safety is always a balance of risks and benefits. Some types of cancer have a prognosis of only a few months. Hence the risk-benefit analysis cannot be discussed for drug combinations in general, but depends on the type of disease, the type of drugs involved and the condition and informed choice of each patient.

Drug interactions are a known cause of adverse events, but, given that multi-therapy is common and essential for many patients (most hospitalized patients receive at least six drugs [30]), it is preferable to develop formal methods of assessment, as we suggest, rather than leaving the development of multi-therapies to the empirical decision of individual physicians.

*Non-linearities and control landscapes*

We have mentioned in the Introduction that one of the desirable features of these algorithms is the capacity of dealing with non-linearities in drug combinations. The most suitable measures of non-linearity can be obtained by building an n-dimensional "control landscape", where the dimensions are the drugs, at different doses. The notion of landscape represents a commonly used concept in the analysis of many complex systems encountered in physics, biology, computer science and engineering [7]. Several features can provide a quantitative characterization of these landscapes, such as the density of optima [31] or the ruggedness [7]. The ruggedness measures the correlation of the biological score to be optimized in "neighboring" positions and can be obtained by defining random walk processes in the drug configuration space, and by calculating the correlation length of the score in such processes [31]. The landscape could also be modified using system-wide biological data (omic data) to reduce non-linearities. This omic warping is analogous to approaches commonly used in physics.

While the tests in cancer cell lines reported here do add evidence supporting the efficacy of the suggested algorithms, it would be desirable in future experiments to give priority to the collection of fully factorial datasets. Comparisons with random samples have several limitations, including the fact that the true optimum is unknown. Fully factorial datasets are, when experimentally feasible, more informative, allowing the characterization of the landscapes and the evaluation of alternative search algorithms.

*Integrating other information in the algorithms*

There is a more general rationale supporting the use of algorithms integrating information on the state of the system with iterative measurements. The Artificial Intelligence community realized at the beginning of the 90s that robots could not manage a complex environment utilizing only explicit models of reality [32]. An alternative approach that started from simpler stimulus-response algorithms was more successful and was later integrated with the older models in hybrid architectures [32]. The proponents of this approach (among them Rodney Brooks) argued that this process was similar to the evolution of the nervous system, which is based on stimulus-response mechanisms of increasing complexity in lower invertebrates, integrated (but not replaced) by representations of reality within the brain of higher organisms. See also figure 1 of Pfeifer et al [33]. Similarly we can start from "stimulus-response" algorithms and then improve them using progressively more detailed and mechanistic models of biological networks. The algorithms we have described are composed of several iterations, each depending on the previous response of the system. As pointed out [34], control and optimization algorithms do contain information about the system, when effective, but only in an implicit form. This approach, used to control very complex and partially unknown systems by natural evolution and by possibly the most ambitious attempt to emulate evolution, building intelligent machines, is a general strategy that motivated the development of our algorithms.

It is useful to consider how system-wide molecular data (such as genomic, proteomic, metabolomic and transcriptomic data) could be used in the context of our searches. These omic datasets could affect the ranking in two ways: as objects of multivariate analysis and as parameters of mechanistic network models.

Pattern recognition methods and multivariate statistics can be used to analyze system-wide molecular data [35]. With these models, it could be possible to distinguish the groups studied in a multi-dimensional representation. For example, it might be possible to test whether a combination brings the metabolic and transcriptional profiles of treated cells or organisms closer to that of the target state and by how much. A similar approach was used in a recent publication by Lamb et al [36], where a single score was obtained to represent the response of a breast cancer cell line to drugs. The score was a summary of multivariate biological data (microarrays). This statistical approach is justified by the fact that not all molecular information is included in the network models, but is expected to play a lesser role as the comprehensiveness of the models improves.

Metabolic models similar to that described in our recent paper on *Drosophila* hypoxia may also play a role [37]. Gene expression data of metabolic enzymes and NMR measurements of metabolites for individual treatments could be added to the model and the effect of combining the interventions can be simulated. The model can provide summary measures that have an important effect on function, such as ATP production, and are ideally suited as weighted modifiers of the algorithm rankings. For the cancer experiments we could iteratively modify the apoptosis computational networks described in our recent paper [1]. To reflect the results of intervention experiments, one could add to the model the targets of all the drugs used, and use microarray data specific for the cell types to modify the simulations. As our biological knowledge improves, mechanistic models should play an increasing role.

The algorithms described here are suitable as frameworks to integrate imperfect information from different sources. The information can be used to modify the rankings and fully factorial datasets can be used to assign weights to different types of information. For example, if the cytoprotective protein Bcl-2 is overexpressed in a target cell type or if network simulations indicate that it is an important control node, one could modify the ranking metric of combinations including drugs acting on it and test whether this improves the efficiency of the algorithms within our fully factorial datasets.

*Potential applications to personalized medicine*

There is great interest in personalized medicine and it is clear that personalized therapy requires combinations, since we cannot develop a different drug for each patient. The information on the state of the system that we suggest should be incorporated in the algorithms, can at the same time provide a molecular profile corresponding to each effective combination. In other words an omic-combination dictionary could be built listing the untreated genomic, transcriptomic, proteomic and metabolomic profile optimally responding to a drug combination, and this information could guide therapy in individual patients.

The algorithms could be used not only to find optimal combinations for specific diseases but also for individual patients when repeated sampling is feasible, for example in studies of chemosensitivity of cells from the blood of leukemia patients [38].

*Conclusion*

Novel technology for high-throughput screening and for omic data measurements might allow us to develop new combined pharmacological interventions adapting algorithmic and theoretical approaches from more quantitative sciences

We report data from computational simulations and from biological experiments in vivo and in cell culture, suggesting that modified search algorithms from information theory have the potential to enhance the discovery of novel optimal or near-optimal therapeutic combinations.

It would be desirable to obtain a larger number of fully factorial datasets, for different biological systems. This would allow a direct comparison of the algorithms reported here with other reasonable alternatives, such as stochastic algorithms. Fully factorial datasets would be even more useful if they were to include system-wide molecular (omic) data, at least for the single drug and for untreated cases. While this might require a considerable experimental effort, it would allow this area of research to be firmly established and provide a resource for scientists with different algorithmic backgrounds to test their ideas.

Several colleagues have already pointed out analogies with other computational problems within their fields of expertise that might lead to useful alternative approaches. For example a colleague has suggested that exploring alternatives within the class of "online algorithms" is a promising area of future work. Other colleagues have proposed that modern biologically-inspired heuristic methods, such as "particle swarm optimization", might also be used to search for optimized drug combinations. In the next few years we plan to obtain and make fully available on the web additional fully factorial datasets for drug-induced selective cell death, and we hope that this will stimulate interdisciplinary interest in this approach to the problem of multi-drug therapy.

**MATERIALS AND METHODS**

*Drosophila physiology*

A detailed account of the *Drosophila* cardiac aging model was presented previously, in which an age-dependent decline in Drosophila cardiac rate under stress was reported [10]. We developed new methods for imaging rapidly and non-invasively the adult *Drosophila* heart and for automated measurement of heart rate and its variability.

To assess exercise capacity in *Drosophila* and changes with age, climbing velocity was measured using a method described by Gargano et al. [39], modified to include image processing that allowed individual flies to be studied.

The flies were transferred into 15-ml tubes and the operator tapped the top of the tube. Owing to their capacity for geotaxis orientation, flies tend to climb upwards. A digital imaging system/camera (Motionscope PCI, Redlake Imaging MASD, Inc.) with an attached Vivitar wide-angle lens, was used to capture video sequences at 60 frames per second of the flies as they climbed the tube. Images were analyzed with software (MotionScope 2.21.1) and for each fly within the tube an individual velocity was obtained.

The selected compounds and doses (in the fly food) were: doxycycline, with concentrations at 0.5 mg/mL and 1 mg/mL; sodium selenite, at 0.005 mg/mL and 0.0125 mg/mL; zinc sulfate, at 0.5 mg/mL and 1 mg/mL; and resveratrol, at 0.25 mM and 0.5 mM.

*Computational Simulations*

The data sets, used to test the SS' and SS-TD' algorithms, were created using the apoptosis model [1], with some changes concerning the search procedure and the output value. Instead of performing an exhaustive search on all the nodes of the network, we limited the search to a randomly chosen subgroup of nodes. We also used as output value the difference of the cubic value of one individual compared to the average of the cubic sum of the remaining population, to reward the individuals with the highest values.

Confirmatory simulations were also performed, to test the robustness of our findings by changing several parameters. The parameters were the number of states for nodes and links, the starting values for the states, the ranges of the output of the simulation, and the nodes selected for the interventions.

The software was written in C++ and implemented on 32-nodes of a 64-bit Linux cluster with 2GB of memory per node. The longest searches required about 30 minutes of computation.

The analysis of the collected data consisted of three separated steps: sort, search algorithm and statistical analysis. For the first step, a quick sort implementation was used creating different ranks for each individual. In the second step, all the algorithms and random execution returned information for each rank. These were used in the last step, where we collected the statistical analysis data, dividing the resulting population into different samples, to compare each algorithm with the others.

Owing to the dimension of the data, it was necessary to limit the number of analyzed nodes to a maximum of 9. Computational time was significant only for the sorter, requiring several hours for the largest files on an entry-level Linux workstation .

*Cancers cells*

ATP is a marker for cell viability because it is present in all metabolically active cells and the concentration declines very rapidly when the cells undergo necrosis or apoptosis. Human tumor cells DOHH2 and RS11846 were maintained as suspension cultures at standard conditions: humidified atmosphere with 5% carbon dioxide, at 37°C in an incubator, using RPMI-1640 medium, supplemented with 10% heat-inactivated fetal calf serum and 2 mM L-glutamine. Cells were kept in log phase via replacement of cellular suspension aliquots by fresh medium two or three times weekly. Stock solutions of the 6 chosen drugs were freshly prepared in water (Vincristine), physiological saline solution (Rituximab) or DMSO (Etoposide, Q-VD-Oph, Apogossypol and Dexamethasone). The stock solutions were diluted with RPMI-1640 in order to obtain the desired final concentrations. Less than 0.5% of the solvent was present at the final dilutions. All the procedures related to cell culture, drug preparation, and treatment were carried out in a laminar flow cabinet.

Briefly, exponentially growing cells were seeded in 96-well plates (90 $\mu$L aliquots/well) at a density of 5.55 $10^4$ cells/mL and 10 $\mu$L of drug solution were added. Final concentrations of the drug were the following: Vincristine (0.01, 0.1, or 0.5 nM), Etoposide (0.01, 0.1, or 1 $\mu$M), Apogossypol (1, 2.5, or 4 $\mu$M), Q-VD-OPh (5, 10, or 25 $\mu$M), Rituximab (5, 15, or 20 $\mu$g/mL), Dexamethasone (0.1, 1, or 25 $\mu$M). Plates were incubated for 60 hours. After the incubation, 30 $\mu$L aliquots of ATPlite reconstituted reagent (Perkin-Elmer) were added to every well. The plates were shaken for 3 minutes at 750 rpm (Eppendorf MixMate). The absorption of the samples was measured using a monolight 3096 microplate luminometer (BD). Ten $\mu$L of a 10 mM ATP solution was added to every well as internal standard. The plates were shaken for 2 minutes at 750 rpm and read.

Selectivity was defined as the difference in % survival between the two cell types.

*Statistical Analysis*

All results are expressed as mean ± standard error of the mean. For comparisons of 2 groups unpaired t tests were used (non-parametric tests were also significant) and for comparison of more than 2 groups we used one-way analysis of variance with Bonferroni correction for post-test comparisons. The Drosophila data presented in Figure 1 were analyzed using the chi square test for trends and results were confirmed using one-way analysis of variance with linear test for trend. The number of combinations in the introduction was obtained using Newton's Binomial series up to the $6^{th}$ order. The statistical software used was Prism (GraphPad).

**ACKNOWLEDGEMENTS**

We wish to thank Andrew Viterbi for suggesting the use of the stack sequential algorithm in the search for optimal combination therapies and for numerous helpful discussions. We also wish to thank Shinichi Kitada for providing cell lines and reagents for the cell culture experiments, Amarnath Gupta for contributing to improve the description of the algorithm, Phillip Duxbury for general discussions, Carlo Piermarocchi for general discussions and for the *Drosophila* landscape calculations, Sybille Sauter for many useful comments.

**FIGURE LEGENDS**

**FIGURE 1 – The Drosophila fully factorial dataset.** The number on the right of each combination is a summary score (z-score) obtained from the three phenotypes measured in aged flies: maximal heart rate, exercise capacity and survival. Scores are ordered in descending order with the best on top. The 4 columns show, from left to right, the effects of using 1, 2, 3 and 4 drugs in combination. The effects do not appear to be additive but complex interactions are present. Larger combinations have significantly larger z-scores.

**FIGURE 2 – Tree representation of the data.** Letters indicate drugs and numbers indicate different doses of each drug. The root (level 0) is the control (no drugs), level 1 is composed of individual drug measurements (singles), level 2 is composed of combinations of two drugs (couples) and so on. The level corresponds to the size of the combination. Both this tree and the tree of Jelinek [13] contain repetitions.

**FIGURE 3 – Trellis-like representation of the data** The data can also be represented as a trellis-like structure, without repetitions. The two data representations shown in Figures 2 and 3 (tree and trellis-like, respectively) correspond to the alternative data representations used for coding algorithms [14]. Figure 3 should also be compared to Figure 1, showing the Drosophila dataset. In the more complex Figure 1 the trellis is oriented from left to right and the edges (lines) are not shown directly (for simplicity) but should be seen as connecting each combination with those on the next level that contain the same set of colors. Each color indicates a different drug-dose pair. Additionally, Figure 1 shows ordering according to the summary score (z-score).

**FiGURE 4 – The three classes of algorithms** Three types of strategies for searching the data tree. In our case the starting point was an exhaustive set of measurements of all the single drugs and doses we selected. It is also possible to start the tree search at a higher level, for example after having tested all the couples.

**FIGURE 5 – SS' pseudocode.**

**FIGURE 6 – TD' pseudocode.**

**FIGURE 7 – SS-TD' pseudocode.**

**FIGURE 8 – Simulations using 6 to 9 drugs.** The 3 approaches described (randomly chosen group of combinations, SS' algorithm and the SS-TD' algorithm) are compared. We report the % of tests (average ± SEM from 75 simulations) that can find the best combination, for interventions using 6 to 9 drugs. The cost is expressed as the number of tests. The decline in success rate with an increasing number of drugs for the random and SS' groups is probably explained by the decreasing proportion of the total possible combinations tested (shown as decreasing ratio). This is not the case for SS-TD', but we do not yet know if this is a general property of the algorithm. The total number of possible combinations increases exponentially with the number of drugs and becomes quickly too large for our biological measurements (for example with 9 drugs the total is 262144, see rightmost column) and therefore justifies the necessity of an algorithm to limit the experimental space. The maximal cost for the two algorithms (bottom line) is still within the reach of many experiments. This simulation was done using 3 doses per drug.

**FIGURE 9** – **Cancer cell experiments and the SS-TD' algorithm.** The colors indicate the selectivity of the drug interventions and the aim is to find treatments with high selectivity for one of the cell lines, shown as dark blue. The red shades are partially selective for the other cell line. After measuring the effects of individual drugs, shown as equally distributed in their effects on the two cell lines (Single column), we follow the steps of the SS-TD' algorithm on top and compare it with random testing of combinations of the same size in the lower part of the figure. A statistically significant enrichment of the desired selective combinations is shown.

**FIGURE 10** – **The most general algorithmic approach.** The loop of biological testing and ranking indicates the algorithms described in detail in section 2.2 of the Results. The mechanistic models are examples from our recent publications [1,37] and the system-wide molecular data (omic data), to be collected for individual interventions, represent microarrays and NMR metabolomics. The arrows indicate a flux of information from the system wide molecular data. The network or pathway models and the rankings are incorporating this information but they are not uniquely determined by it. The models are also built using legacy data from the literature and the rankings are produced by the algorithms described in the Results.

Search algorithms for drug combinations

**FIGURE 1**

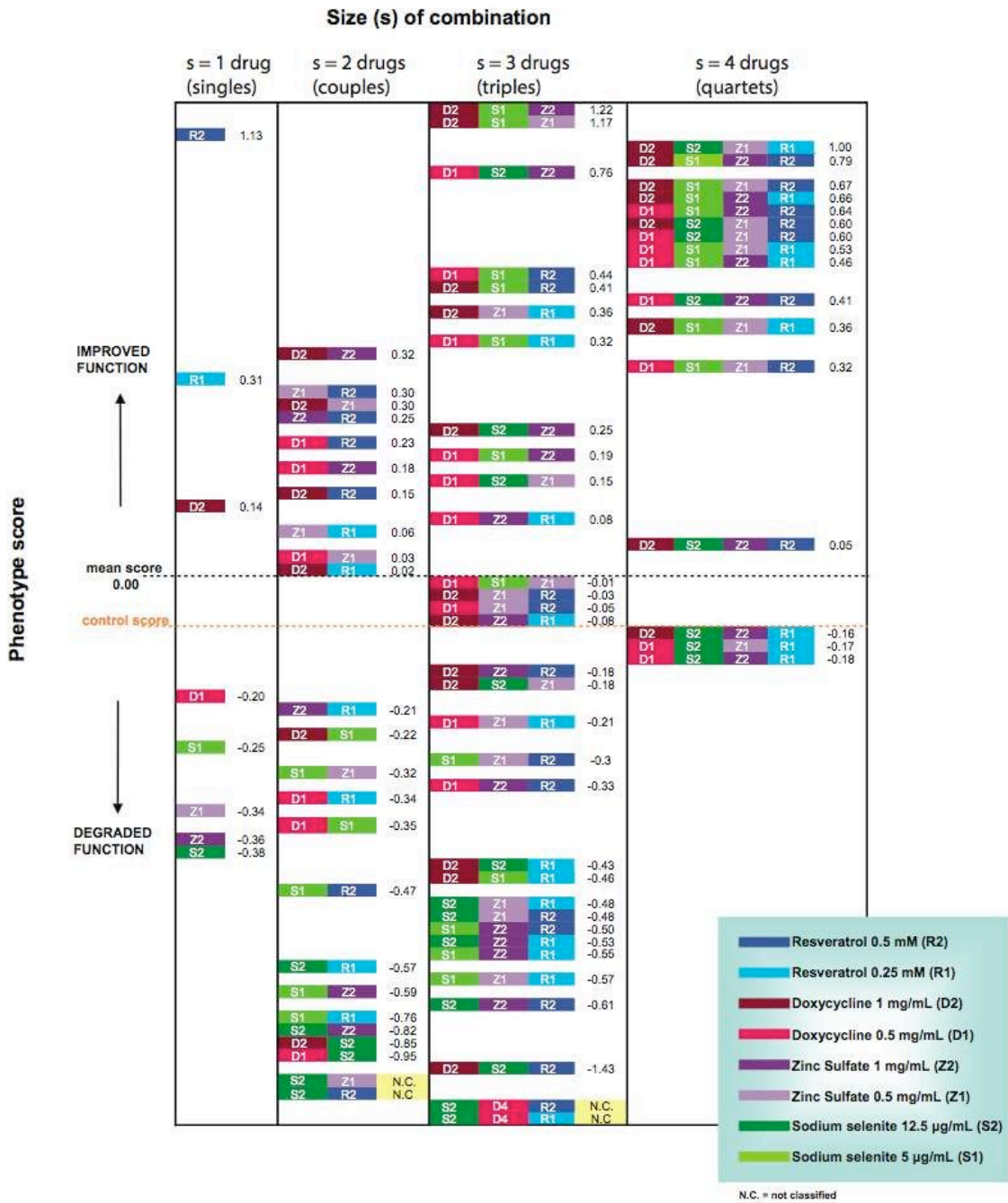

**FIGURE 2**

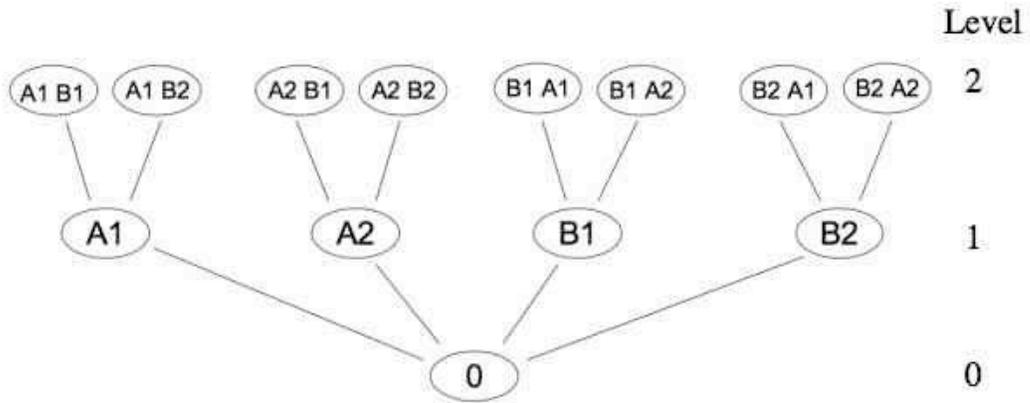

**FIGURE 3**

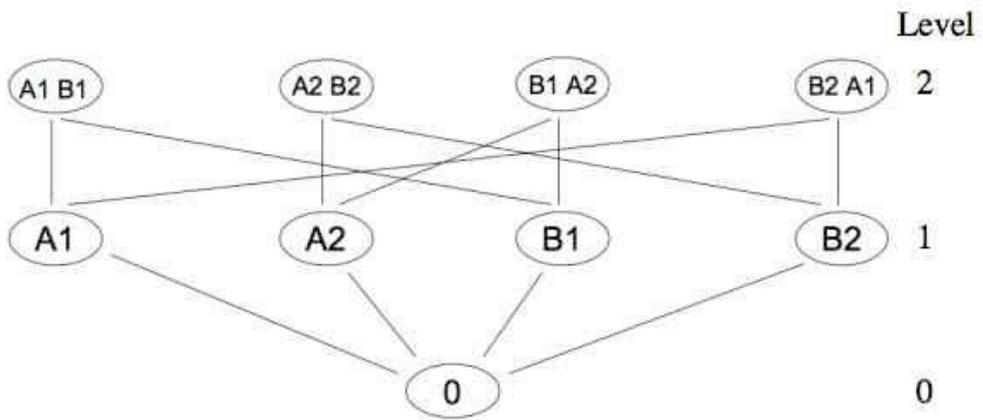

**FIGURE 4**

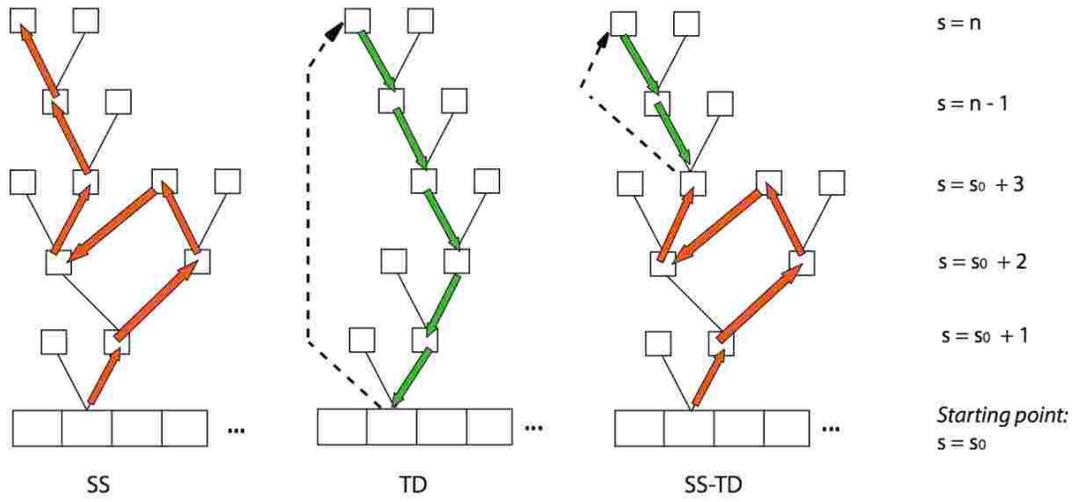

Search algorithms for drug combinations

**FIGURE 5**

```
SS' pseudo code

S1 ┌ FOR all D in DRUGS                          //Measure the drugs individually at all doses (singles)
   │     FOR all d in DOSES
   │         STORE (D,d) in C
   │         STORE Score(C) in C_LIST₁
   │     END FOR
   │ END FOR
   │ SORT C_LIST₁
   └ COPY C_LIST₁ to C_LIST_summary

S2 ┌ FOR all D in C_LIST₁                        // Keep only the best dose for the single
   │     DELETE from C_LIST₁ all d EXCEPT the best score   // drugs
   └ END FOR

S3   SET C_best = C_LIST₁ [best score]           // Find the best single drug

     SET n_tries = number of backtracks allowed

S7 ┌ FOR all combination sizes [2,…,n]           // Move from couples to the maximum size
   │     SET iteration = 1
   │ S6  WHILE (iteration <= n_tries)            // The number of backtracks is counted
   │ S4      ┌ FOR all D in DRUGS EXCEPT those in C_best
   │         │     FOR all d in DOSES            // Add each unused drug to the current best
   │         │         COMBINE (C_best, D, d) as C   // combination, moving one level up the tree
   │         │         STORE Score(C) in C_LIST_size
   │         │     END FOR
   │         │ END FOR
   │         │ SORT C_LIST_size
   │         └ COPY C_LIST_size to C_LIST_summary

   │ S5      ┌ IF (C_LIST_size [best score] >= C_best) THEN   // If any of the new combinations improves on
   │         │     SET C_best = C_LIST_size [best score];     // the current best, extend from there; if not,
   │         │     BREAK FROM WHILE;                          // backtrack and choose the next best
   │         │
   │         │ ELSE
   │         │     INCREASE iteration of 1;
   │         │     SET C_best = C_LIST_size-1 [iteration]
   │         └ END IF

   │     END WHILE
   │ END FOR
   └ SORT C_LIST_summary                         // At the end of the algorithm we obtain a ranked
                                                  // summary list of all the combinations tested
```

**FIGURE 6**

TD' pseudocode

```
S1  FOR all D in DRUGS                              //Measure the drugs individually at all doses (singles)
        FOR all d in DOSES
            STORE (D,d) in C
            STORE Score(C) in C_LIST_1
        END FOR
    END FOR
    SORT C_LIST_1
    COPY C_LIST_1 to C_LIST_summary

S2  SET C_best = C_LIST_1 [best score]              // Find the best single drug

S3  SET size = n
    FOR all possible C of (C_best, all other D in DRUGS, all d in DOSES) with |C|=size
        STORE Score(C) in C_LIST_n
    END FOR                                         // Add all unused drugs (at all doses) to the current best
    SORT C_LIST_n                                   // single drug, moving to the top of the tree
    COPY C_LIST_n to C_LIST_summary

S8  WHILE size > 1                                  // Move from the maximum size to couples
    S4  CALCULATE median of C_LIST_size;
        DELETE from C_LIST_size all C below the median   // Select combinations scoring in the top half

    S5  COUNT (D,d) occurrences in C_LIST_size      // Count how often each drug and dose appear in
        STORE (D,d) occurrences in Dd_LIST          // the top half

    S6  FOR all D in Dd_LIST
            DELETE from Dd_LIST all (D,d) EXCEPT for the most frequent d   // Select only the most frequent
        END FOR                                                            // dose for each drug

    S7  DECREASE size by 1                          // Use this selection to create all possible
        FOR all possible C of (D,d) in Dd_LIST with |C|=size   // combination of the next smaller size,
            STORE Score(C) in C_LIST_size           // moving one level down the tree.
        END FOR
        SORT C_LIST_size
        COPY C_LIST_size to C_LIST_summary          // At the end of the algorithm we obtain a ranked
                                                    // summary list of all the combinations tested
```

**FIGURE 7**

SS-TD' pseudo code

```
CALL Algorithm SS' until size = J           // Follow algorithm SS' until combinations of size J are reached,
CALL Algorithm TD' from S3 until size = J+1 // from there jump to the largest combinations and follow
                                            // algorithm TD' for the remaining levels
```

Search algorithms for drug combinations

**FIGURE 8**

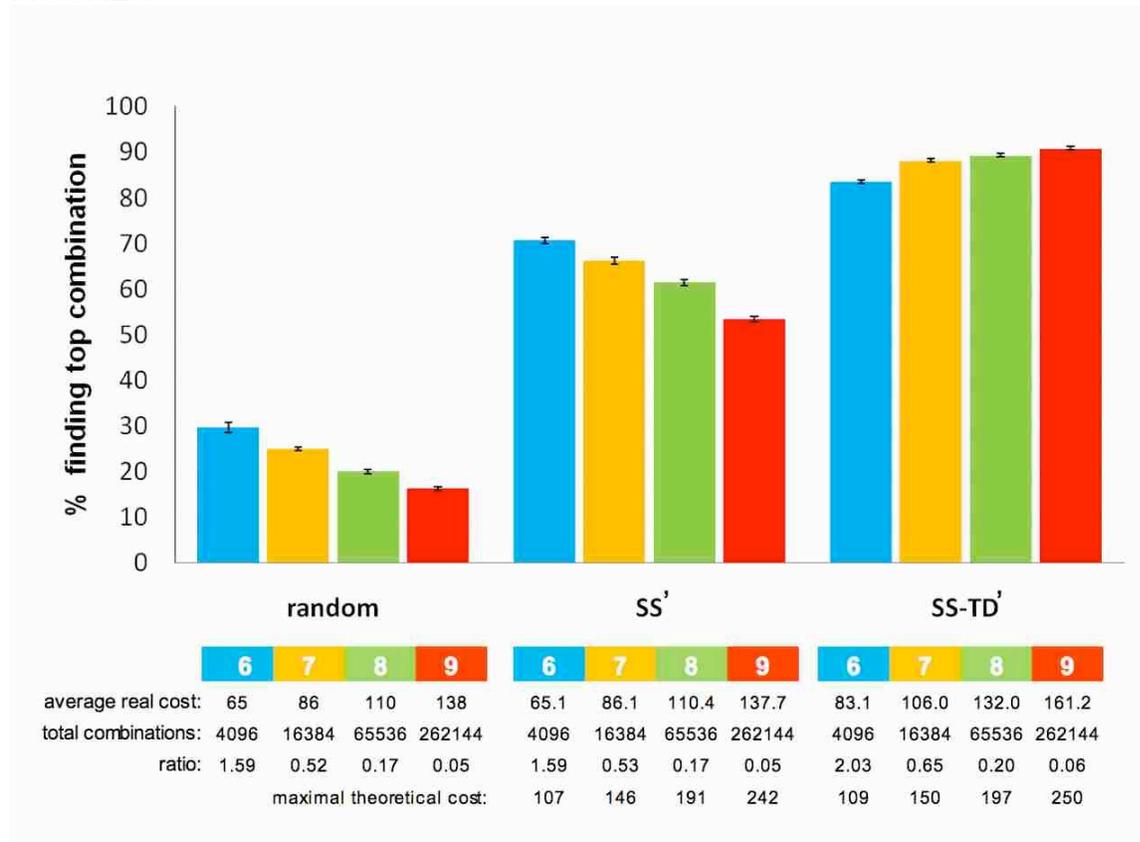

Search algorithms for drug combinations

**FIGURE 9**

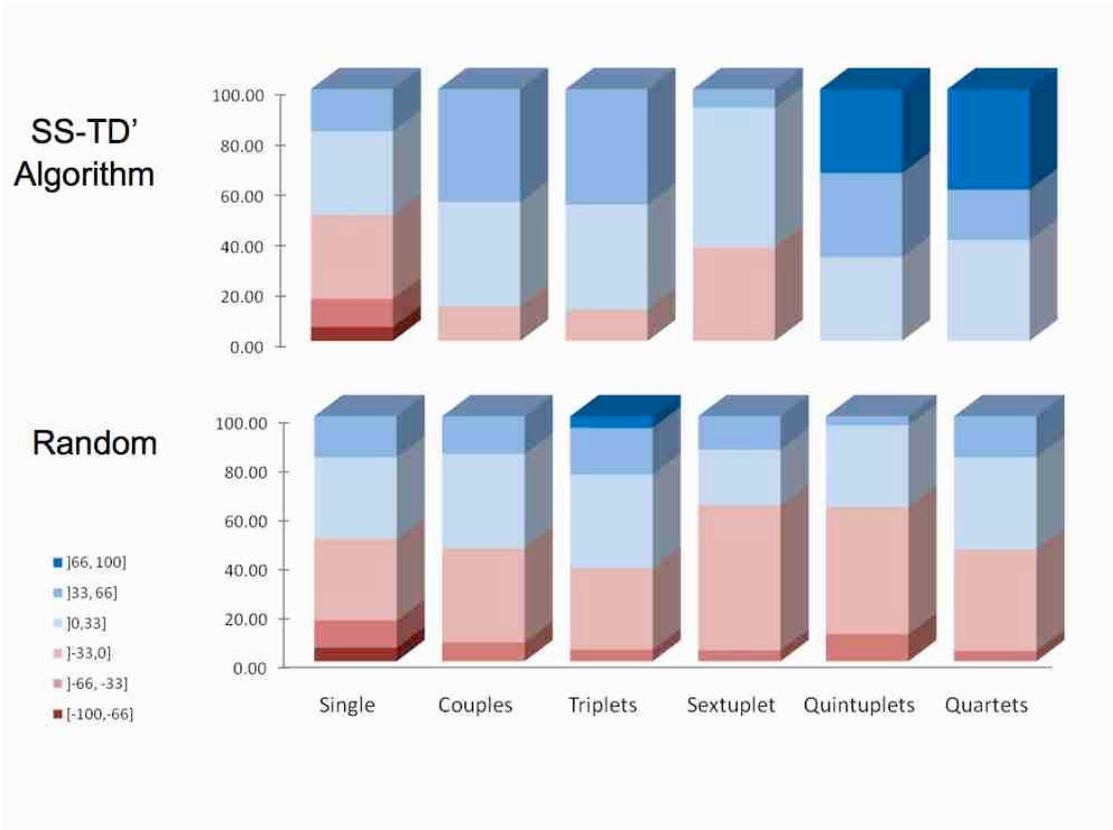

FIGURE 10

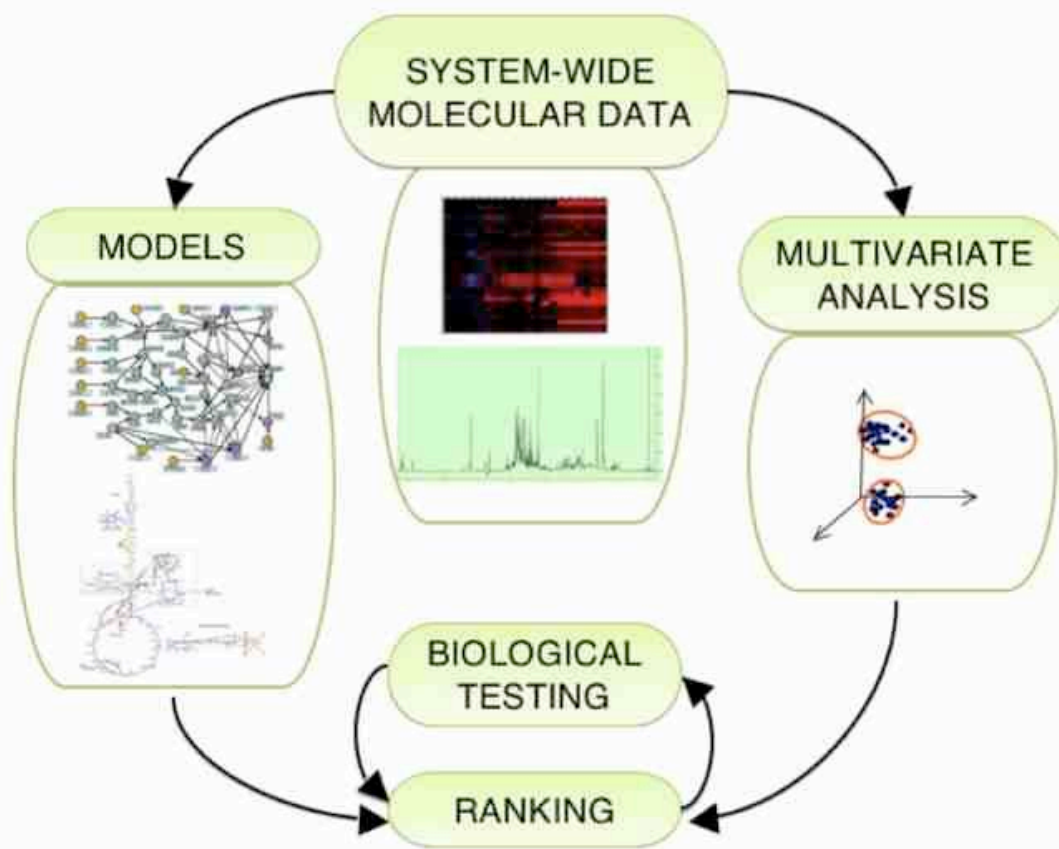

Search algorithms for drug combinations

**SUPPLEMENTARY MATERIAL**

**TABLE S1**

**Number of possible combinations of a given size (s), for two doses (m = 2)**

|  | Size of combination (s) | | | | | | | | | |
|---|---|---|---|---|---|---|---|---|---|---|
| Number of drugs (n) | 0 | 1 | 2 | 3 | 4 | 5 | 6 | 7 | 8 | 9 | Total = $(m+1)^n$ |
| 0 | 1 | | | | | | | | | | 1 |
| 1 | 1 | 2 | | | | | | | | | 3 |
| 2 | 1 | 4 | 4 | | | | | | | | 9 |
| 3 | 1 | 6 | 12 | 8 | | | | | | | 27 |
| 4 | 1 | 8 | 24 | 32 | 16 | | | | | | 81 |
| 5 | 1 | 10 | 40 | 80 | 80 | 32 | | | | | 243 |
| 6 | 1 | 12 | 60 | 160 | 240 | 192 | 64 | | | | 729 |
| 7 | 1 | 14 | 84 | 280 | 560 | 672 | 448 | 128 | | | 2187 |
| 8 | 1 | 16 | 112 | 448 | 1120 | 1792 | 1792 | 1024 | 256 | | 6561 |
| 9 | 1 | 18 | 144 | 672 | 2016 | 4032 | 5376 | 4608 | 2304 | 512 | 19683 |

The number of possible combinations of size $s$, chosen from a set of $n$ drugs over $m$ doses, can be calculated by the formula

$$_n^sC * m^s = \frac{n!}{s!(n-s)!} * m^s$$

The distribution of this number for a given $n$ reaches a maximum at intermediate sizes, and declines as $s$ approaches $n$. The reason for this stems from the formula for combinations, which is based on the binomial distribution. The total number of combinations of all sizes for $n$ drugs over $m$ doses in this fully factorial dataset is $(m+1)^n$.

**TABLE S2**

Example of level-by-level cost comparison:
Fully Factorial (FF) vs SS'

Example parameters:
  Number of drugs (n) = 6
  Number of doses (m) = 3

| Tree level | Number of Measurements FF | SS' Best case | SS' Worst case |
|---|---|---|---|
| 0 | 1 | 1 | 1 |
| 1 | 18 | 18 | 18 |
| 2 | 135 | 15 | 29 |
| 3 | 540 | 12 | 24 |
| 4 | 1215 | 9 | 18 |
| 5 | 1458 | 6 | 12 |
| 6 | 729 | 3 | 6 |

In the best case, the SS' algorithm never backtracks and can be shown to require a total of $\left(\frac{mn}{2}\right)(n+1)$ measurements, over $n$ parallel measurement cycles. On average, the algorithm performed approximately 2 measurements worse than this best-case scenario for all $n$ in simulations. The worst-case complexity involves a maximum number of backtracks (2 in our implementation) at each level in the tree. This increases the total measurements to $mn(n+1) - mn - 1$, essentially doubling the best-case measurements. In the worst case there are $2n - 2$ parallel measurement cycles.

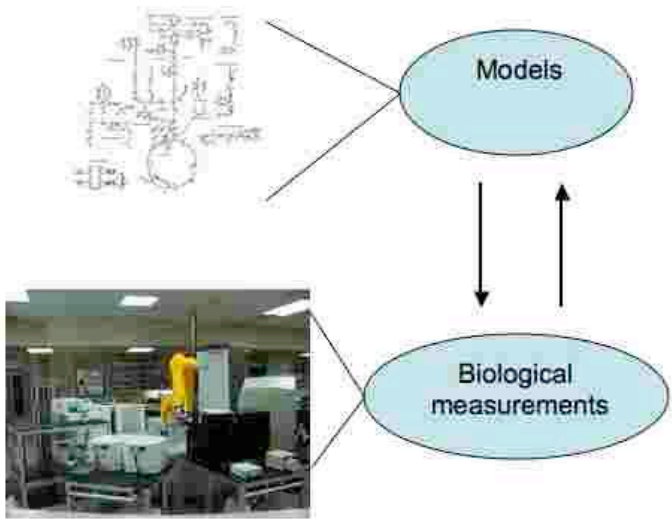

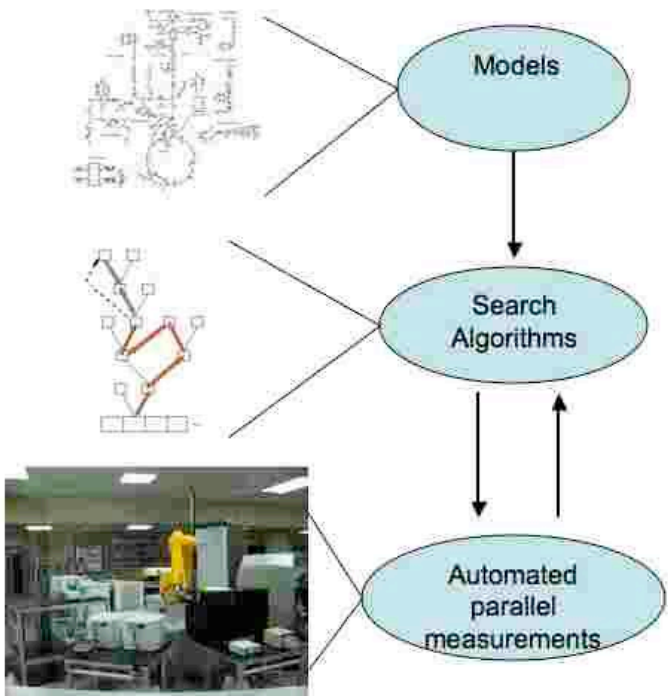

**FIGURE S1 – Comparison of our approach with classical systems biology.**
The upper panel shows the classical approach to systems biology as a cycle of quantitative modeling and biological measurements. The lower panel shows the essential elements of the control-oriented approach we suggest. The middle element of the lower panel represents iterative algorithms that do not contain an explicit model of the reality to be controlled. The approach is supported by the appreciation of a fundamental limit of quantitative biological models (upper element of both panels), which are useful, indeed probably essential, but not sufficient for effective control. This limit is not only due to the complexity of biological organisms but also to their variation, which is one of the essential components of the process of natural selection, and therefore an unavoidable distinctive feature of biological systems.